\documentclass[reprint, amsmath, amssymb, aps, superscriptaddress, prb]{revtex4-1}
\usepackage{graphicx} 
\usepackage{dcolumn}
\usepackage{bm}
\usepackage{color}
\usepackage{hyperref}

\begin{document}
\title{Anomalous Fano factor as a signature of Bogoliubov Fermi surfaces}
\author{Sayan Banerjee}
\thanks{These authors contributed equally to this work.}
\affiliation{Max-Planck-Institut f\"ur Festk\"orperforschung, Heisenbergstrasse 1, D-70569 Stuttgart, Germany}
\affiliation{Institute for Theoretical Physics III, University of Stuttgart, D-70550 Stuttgart, Germany}
\author{Satoshi Ikegaya}
\thanks{These authors contributed equally to this work.}
\affiliation{Department of Applied Physics, Nagoya University, Nagoya 464-8603, Japan}
\author{Andreas P. Schnyder}
\email{a.schnyder@fkf.mpg.de}
\affiliation{Max-Planck-Institut f\"ur Festk\"orperforschung, Heisenbergstrasse 1, D-70569 Stuttgart, Germany}
\date{\today}
\begin{abstract}
Noise spectroscopy is a key technique to investigate the nature and dynamics of charge carriers in superconductors.
The recently discovered superconducting hybrids with Bogoliubov Fermi surfaces exhibit a particularly 
intriguing and rich charge dynamics, as their charge carriers consist of both Cooper pairs and an extensive number of Bogoliubov quasiparticles.
Motivated by this, we compute the noise spectra of Bogoliubov Fermi surfaces and identify their key signatures
in the differential conductance and the Fano factor. 
Specifically, we consider a semiconductor/superconductor hybrid device with an  in-plane magnetic field, which exhibits several Bogoliubov Fermi surfaces.
The number and orientation of  the Bogoliubov Fermi surfaces in this device can be readily controlled by the applied magnetic field, which in turn alters
the noise signal.
In particular, we find that the Fano factor exhibits a reduced value, substantially lower than two, 
whenever the charge dynamics is governed by a large number of Bogoliubov quasiparticles.
 Using experimentally relevant parameters, we make a number of specific predictions for the noise spectra, that can be used as direct evidence of Bogoliubov Fermi surfaces.
In particular, we find that the Fano factor as a function of magnetic field and spin-orbit coupling exhibits characteristic discontinuities at the transition lines that separate phases
with different number of Bogoliubov Fermi surfaces.
\end{abstract}
\maketitle

\section{Introduction}
Since the discovery of unconventional superconductors in the 1980s \cite{bednorz_muller_1986,heavy_fermion_RMP,keimer_review_nat_15}, the
study of low-energy Bogoliubov quasiparticles (BQPs)
has been an important research topic
with both fundamental and applied interest \cite{handbook_SC_2007}.
Usually, unconventional superconductors exhibit only point nodes or line nodes in their pairing gap,
with dimension strictly smaller than the normal-state Fermi surface,
due to the loss of condensation energy. 
However, recent findings have shown that there can also exist two-dimensional nodes in the superconducting gap, which form a Fermi surface of BQPs
whose dimension is equal to the normal-state Fermi surface.
For instance,  Bogoliubov Fermi surfaces (BFSs) can exist in multi-band superconductors with broken time-reversal symmetry~\cite{volovik_zeroes_1989,agterberg_bogoliubov_2017-1,timm_inflated_2017-1,brydon_bogoliubov_2018-1,menke_bogoliubov_2019}, 
in nodal superconductors with external Zeeman potentials~\cite{yang_response_1998,gubankova_breached_2005,setty_bogoliubov_2020},
and in superconductors carrying large supercurrents~\cite{fulde_tunneling_1965,zhu_discovery_2021}.
In addition, BFSs can be created in superconducting hybrids, such as 
semiconductor-superconductor heterostructures with strong spin-orbit coupling (SOC) and an external Zeeman field~\cite{phan_detecting_2022,yuan_zeeman-induced_2018}.
The recent experimental observation of BFSs in superconductor hybrids~\cite{phan_detecting_2022,zhu_discovery_2021} has further stimulated this field.

The presence of BFSs with their extensive number of BQPs leads to a large density of states at zero energy, 
which can be detected by a number of experimental probes. 
For example, the large zero-energy density of states can be directly observed in   
tunneling conductance~\cite{setty_bogoliubov_2020,lapp_experimental_2020} and specific heat~\cite{setty_bogoliubov_2020,lapp_experimental_2020}, and moreover 
leads to characteristic signatures in quasiparticle interference patterns~\cite{zhu_discovery_2021}.
While these experimental probes detect the presence of zero-energy quasiparticles, they do not provide information
about their nature and dynamics. This is where noise spectroscopy comes into play, which can measure
the fundamental charge of the quasiparticles as well as their dynamics. 
For example,  the current-noise ratio (i.e., Fano factor) in the tunneling limit of a fully gapped
superconductor is twice as large as in the normal state~\cite{de_jong_doubled_1994,anantram_current_1996}, indicating
that the charge carriers are Cooper pairs composed of two electrons. 
However, in systems with BFSs, where the charge carriers consist of both Cooper pairs
and BQPs, the Fano factor is expected to be smaller than two, since the effective charge of the BQPs is substantially smaller than that
of Cooper pairs. Hence, noise spectroscopy can give valuable information about the presence of BFSs and can tell us to which degree
BQPs contribute to charge transport.

Motivated by these considerations, we investigate in this Letter the Fano factor of
a semiconductor/superconductor hybrid device which exhibits BFSs that can be controlled
by an external magnetic field. 
After defining the model Hamiltonian of the considered device (Sec.~\ref{sec_II}), we determine in Sec.~\ref{sec_III}
the regions in parameter space of SOC and field strength,
where BFSs exist.  
Interestingly, we find that with increasing field strength our device can be
tuned through two phase transitions, where the number of BFSs increases from zero to two to four (Fig.~\ref{fig:figure2}).
In Sec.~\ref{sec_IV} we present our results on the differential conduction and noise spectroscopy.
We observe that both the differential conductance and the Fano factor exhibit characteristic discontinuities 
at the phase transition lines. In the tunneling limit the value of the Fano factor reduces from two to about one, 
indicating that a substantial part of the charge is carried by BQPs rather than Cooper pairs. 
These results provide key fingerprints for the 
experimental identification of BFSs in the considered device.

\section{Model definition} \label{sec_II}

 %------------------------------------------------------------------------------
\begin{figure}[t!]
\begin{center}
\includegraphics[width=0.475\textwidth]{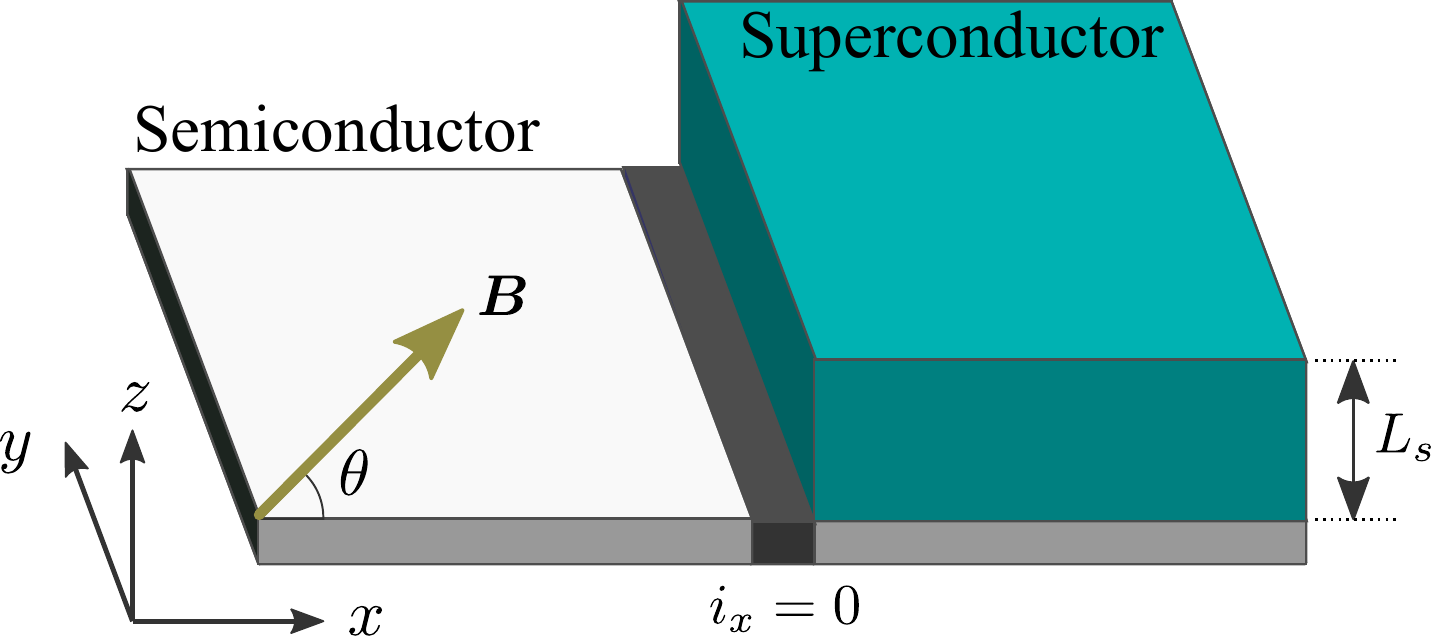}
\caption{Schematics of the considered semiconductor/superconductor hybrid device that exhibits Bogoliubov Fermi surfaces.
An external field $\boldsymbol{B}$ is applied in-plane and can be rotated by the angle $\theta$.
A potential barrier at $i_x=0$ separates the superconducting from the semiconducting region.}
\label{fig:figure1}
\end{center}
\end{figure}
%------------------------------------------------------------------------------

We consider a thin-film semi\-con\-duc\-tor/super\-con\-duc\-tor hybrid device with an external magnetic field applied in-plane, as illustrated in Fig.~\ref{fig:figure1}.  
The superconductor is assumed to be of conventional nature with spin-singlet $s$-wave pairing, while the semiconductor is non-centrosymmetric with a substantial
Rashba SOC, as realized in, e.g., Al/InAs heterostructures~\cite{phan_detecting_2022,marcus_nat_commun_2016}.
The semiconductor is separated into two regions by a potential barrier at $i_x =0$, whose transparency can be controlled  experimentally  
by applying gate voltages. Due to proximity to the parent superconductor, the semiconductor at $i_x >0$ acquires 
an induced $s$-wave pair potential~\cite{phan_detecting_2022,marcus_nat_commun_2016,fraunhofer_interference_marcus_PRB_17,marcus_Phys_Rev_Applied_17,mayer_shabani_APL_19}. As a result, the hybrid device realizes an effective two-dimensional NS junction
with tunable potential barrier and in-plane field. As we will see, the high tunability of the device allows
us to study the conductance and noise signals for different BFSs configurations and junction transparencies.

Mathematically, we describe the considered device by a Bogoliubov-de Gennes
(BdG) Hamiltonian on a three-dimensional cubic lattice with lattice sites $\boldsymbol{r}$ labelled by
$\boldsymbol{r}=i_x\boldsymbol{x}+i_y\boldsymbol{y}+i_z\boldsymbol{z}$,
where the three lattice vectors $\boldsymbol{x}$, $\boldsymbol{y}$, and $\boldsymbol{z}$ are assumed to have unit length $a_0=1$.
We can separate the BdG Hamiltonian into three parts, $H=H_n+H_s+H_i$,
representing the semiconductor, the superconductor, and the coupling between them, respectively. 
The Hamiltonian of the two-dimensional semiconductor at $i_z=0$ is given by
\begin{align}
H_n&=\sum_{\bar{\boldsymbol{r}}}\sum_{\bar{\boldsymbol{w}}=\pm \boldsymbol{x},\pm \boldsymbol{y}}
C^{\dagger}_{\bar{\boldsymbol{r}}+\bar{\boldsymbol{w}}}
\left\{-t_n\hat{\sigma}_0-\frac{i\lambda}{2}\boldsymbol{z}\cdot(\hat{\boldsymbol{\sigma}}\times \bar{\boldsymbol{w}})\right\}C_{\bar{\boldsymbol{r}}}
\nonumber\\
&+\sum_{\bar{\boldsymbol{r}}}C^{\dagger}_{\bar{\boldsymbol{r}}}
\left\{ (\epsilon_n +X_b \delta_{i_x,0}) \hat{\sigma}_0+\boldsymbol{V}_n \cdot \hat{\boldsymbol{\sigma}}\right\}C_{\bar{\boldsymbol{r}}},
\end{align}
where  $\hat{\boldsymbol{\sigma}}=(\hat{\sigma}_x,\hat{\sigma}_y,\hat{\sigma}_z)$
are the three Pauli matrices in spin space, and $\hat{\sigma}_0$ denotes the $(2\times2)$ unit matrix.
Here,
$C_{\bar{\boldsymbol{r}}}=[c_{\bar{\boldsymbol{r}},\uparrow},c_{\bar{\boldsymbol{r}},\downarrow}]^{\mathrm{T}}$
represents a spinor composed of operators $c_{\bar{\boldsymbol{r}},\sigma}$ ($c^{\dagger}_{\bar{\boldsymbol{r}},\sigma}$)
which annihilate (create) electrons with spin~$\sigma$ at  sites $\bar{\boldsymbol{r}}=i_x\boldsymbol{x}+i_y\boldsymbol{y}$ within the semiconductor.
$t_n$ denotes the nearest-neighbor hopping amplitude and $\epsilon_n=4t_n-\mu_n$ is the onsite energy, with $\mu_n$  the chemical potential.
The strength of the Rashba SOC and the potential barrier at $i_x=0$ are denoted by
 $\lambda$ and $X_b$, respectively. 
The in-plane magnetic field $\boldsymbol{B}=B(\cos \theta, \sin \theta, 0)$ induces a 
Zeeman potential $\boldsymbol{V}_n=V_n(\cos \theta, \sin \theta, 0)$  
of strength  $V_n=g_n \mu_B B$, where  $\mu_B$ and $g_n$ deonte the Bohr magneton and the $g$-factor of the semiconductor, respectively.

 The Hamiltonian of the three-dimensional superconductor, located at $i_x>0$ and $0 < i_z \leq L_s$,
 reads
 \begin{align}
H_s=&-t_s \sum_{i_x>0} \sum_{i_y} \sum_{i_z=1}^{L_s}\sum_{\bar{\boldsymbol{w}}=\boldsymbol{x},\boldsymbol{y}}
\left[D^{\dagger}_{\bar{\boldsymbol{r}}+\bar{\boldsymbol{w}},i_z} D_{\bar{\boldsymbol{r}},i_z}+\mathrm{H.c.}\right]\nonumber\\
&-t_s \sum_{i_x>0} \sum_{i_y}\sum_{i_z=1}^{L_s-1}( D^{\dagger}_{\bar{\boldsymbol{r}},i_z+1} D_{\bar{\boldsymbol{r}},i_z} + \mathrm{H.c.})\nonumber\\
&+\sum_{i_x>0} \sum_{i_y} \sum_{i_z=1}^{L_s}D^{\dagger}_{\bar{\boldsymbol{r}},i_z}
\left\{ \epsilon_s\hat{\sigma}_0+\boldsymbol{V}_s \cdot \hat{\boldsymbol{\sigma}}\right\}D_{\bar{\boldsymbol{r}},i_z}\nonumber\\
&+\sum_{i_x>0} \sum_{i_y} \sum_{i_z=1}^{L_s}\Delta 
(d^{\dagger}_{\bar{\boldsymbol{r}},i_z,\uparrow}d^{\dagger}_{\bar{\boldsymbol{r}},i_z,\downarrow} + \mathrm{H.c.}),
\end{align}
where $D_{\bar{\boldsymbol{r}},i_z}=[d_{\bar{\boldsymbol{r}},i_z,\uparrow},d_{\bar{\boldsymbol{r}},i_z,\downarrow}]^{\mathrm{T}}$
is a spinor composed of operators $d_{\bar{\boldsymbol{r}},i_z,\sigma}$ ($d^{\dagger}_{\bar{\boldsymbol{r}},i_z,\sigma}$)
which annihilate (create)   electrons with spin~$\sigma$ at  sites $\boldsymbol{r} = \bar{\boldsymbol{r}}+i_z\boldsymbol{z}$ within the superconductor.
$t_s$ denotes the hopping amplitude in the superconductor, $\Delta$ represents the spin-singlet $s$-wave pair potential,
and $\epsilon_s=6t_s-\mu_s$ is the onsite energy, with $\mu_s$  the chemical potential.
The Zeeman potential in the superconductor is given by $\boldsymbol{V}_s=V_s(\cos \theta, \sin \theta, 0)$,
with strength $V_s = g_s \mu_B B$ and $g$-factor  $g_s$.
Finally, the coupling between the semiconductor and the superconductor is given by
\begin{align}
H_i=-t_i\sum_{i_x>0} \sum_{i_y}( D^{\dagger}_{\bar{\boldsymbol{r}},i_z=0} C_{\bar{\boldsymbol{r}}} + \mathrm{H.c.}),
\end{align}
with amplitude $t_i$. This coupling induces a finite superconducting pair potential in the semiconductor segment $i_x >0$, 
which we compute using lattice Greens function techniques~\cite{lee_anderson_1981,ando_quantum_1991}.

For the numerical calculations, we set the parameters to $t_n=10t_s$, $\mu_n=0.5t_s$, $\mu_s = 3t_s$, $\Delta=0.01t_s$, $t_i=0.2t_s$ and $L_s = 20$.
In addition, we fix the ratio of the $g$-factors to $g_n/g_s=5$ and express the strength of the 
potential barrier by the normalized value  $Z=X_b/(2t_n\sin k_F )$, with $k_F = \arccos[1-(\mu_n/2t_n)]$ the Fermi momentum of the semiconductor. 
We note that the choice of $\Delta$ and $g_n / g_s$ corresponds to the experimentally relevant regime, as observed, e.g., in Al/InAs heterostructures~\cite{phan_detecting_2022,marcus_nat_commun_2016}.   
In the following, we calculate the differential conductance and the Fano factor as a function of junction transparency $Z$, SOC strength $\lambda$, and Zeeman field $V_s$,
for different field orientations $\theta$.
We note that with this choice $V_n$ is given by $V_n =  5 V_s$.

\section{Phase diagram} \label{sec_III}

\begin{figure}[tttt]
\begin{center}
\includegraphics[width=0.5\textwidth]{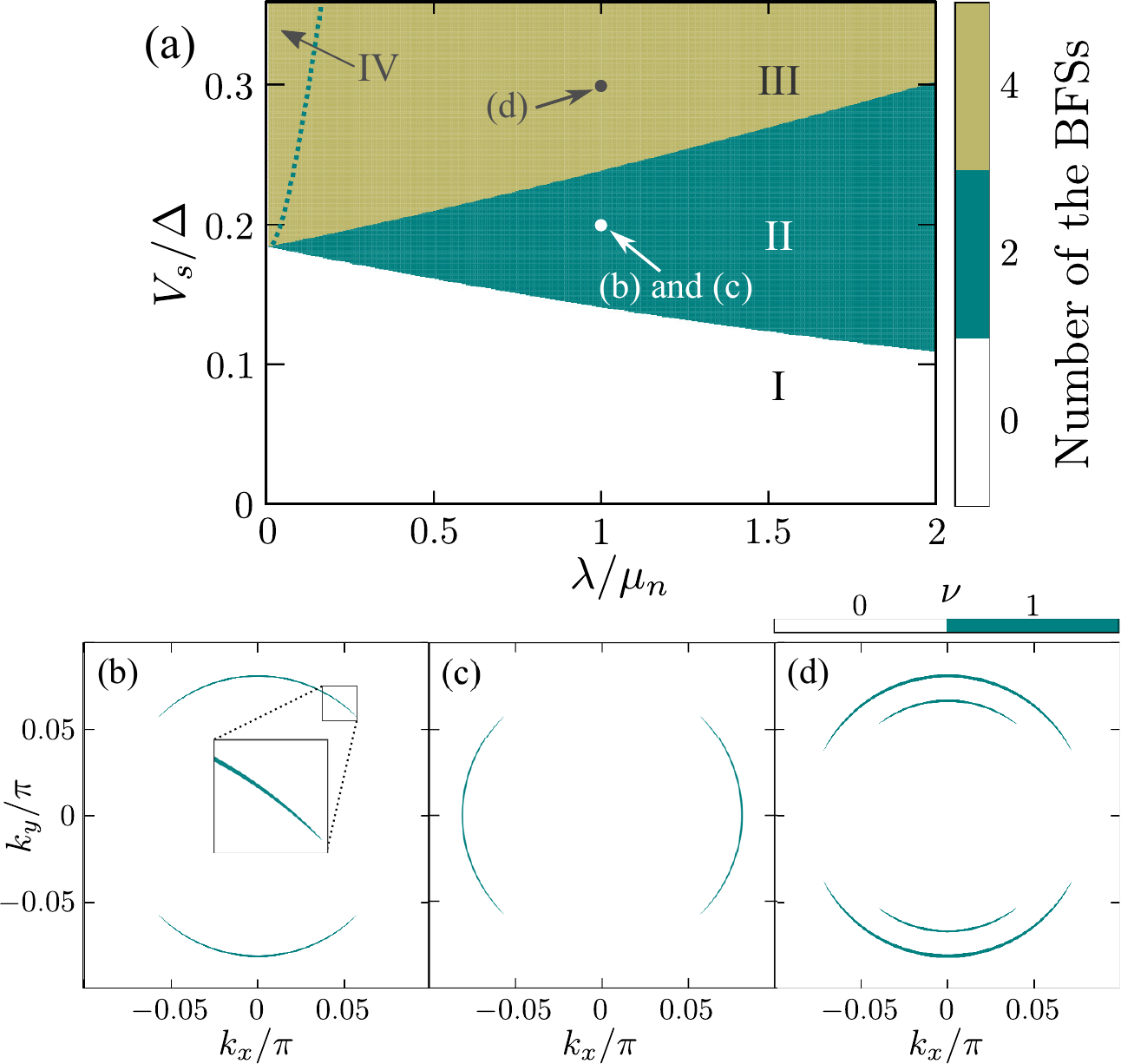}
\caption{
(a) Phase diagram of the device shown in Fig.~\ref{fig:figure1} as a function
of Rashba SOC strength $\lambda$ and Zeeman field $V_s$. 
Phase II has two BFSs, while phases III and IV have four BFSs.
Panels (b), (c), and (d) show the determinant index $\nu$, Eq.~\eqref{eq:index},  
 in the two-dimensional Brillouin zone for
 $(\lambda/\mu_n,V_s/\Delta,\theta/\pi)=(1.0,0.2,0)$, $(1.0,0.2,0.5\pi)$, and $(1.0,0.3,0)$, respectively,
 as indicated in panel (a). 
The colored regions are bounded by BFSs.}
\label{fig:figure2}
\end{center}
\end{figure}

Before calculating the   conductance and noise spectra, we first determine
the parameter regions where BFSs emerge in the considered device.
For this purpose, we remove the semiconductor segment $i_x \leq 0$ and apply  periodic boundary conditions in both $x$- and $y$-directions.
By Fourier transforming the spinors $C_{\bar{\boldsymbol{r}}}$ and $D_{\bar{\boldsymbol{r}},i_z}$, 
as
$C_{\bar{\boldsymbol{r}}} = \sum_{\bar{\boldsymbol{k}}}e^{\bar{\boldsymbol{k}}\cdot \bar{\boldsymbol{r}}}C_{\bar{\boldsymbol{k}}}$
and $D_{\bar{\boldsymbol{r}},i_z}=\sum_{\bar{\boldsymbol{k}}}e^{\bar{\boldsymbol{k}}\cdot \bar{\boldsymbol{r}}}D_{\bar{\boldsymbol{k}},i_z}$,
the BdG Hamiltonian is simplified to
\begin{align} \label{eq_phase_diag_ham}
H=\frac{1}{2}\sum_{\bar{\boldsymbol{k}}} \Psi^{\dagger}_{\bar{\boldsymbol{k}}} H(\bar{\boldsymbol{k}}) \Psi_{\bar{\boldsymbol{k}}},
\end{align}
where  $H(\bar{\boldsymbol{k}})$ is given in the Supplemental Material (SM)~\cite{sm_note} and 
$\Psi_{\bar{\boldsymbol{k}}}=[\mathcal{C}_{\bar{\boldsymbol{k}}},\mathcal{D}_{\bar{\boldsymbol{k}},1}, \cdots \mathcal{D}_{\bar{\boldsymbol{k}},L_s} ]^{\mathrm{T}}$,
with $\mathcal{C}_{\bar{\boldsymbol{k}}}=[C_{\bar{\boldsymbol{k}}},C^{\ast}_{-\bar{\boldsymbol{k}}}]^{\mathrm{T}}$
and  $\mathcal{D}_{\bar{\boldsymbol{k}},i_z}=[D_{\bar{\boldsymbol{k}},i_z},D^{\ast}_{-\bar{\boldsymbol{k}},i_z}]^{\mathrm{T}}$.
In the presence of time-reversal symmetry (TRS) and inversion, the emergence of BFSs can be detected by use of a $\mathbb{Z}_2$ Pfaffian
index, as discussed in Refs.~[\onlinecite{agterberg_bogoliubov_2017-1,timm_inflated_2017-1,brydon_bogoliubov_2018-1,zhao_schnyder_PRL_16}].
In the case of Hamiltonian~\eqref{eq_phase_diag_ham}, however, the Pfaffian index is no longer well defined, since TRS and inversion are broken
by the magnetic field and Rashba SOC, respectively. 
Instead, we introduce for a given wave vector $\bar{\boldsymbol{k}}=(k_x,k_y)$ of the two-dimensional Brillouin zone (BZ)
a $\mathbb{Z}_2$ determinant index $\nu$, which can be used
even in the absence of TRS and inversion. It is defined by
\begin{align}
(-1)^{\nu}=\mathrm{sgn}\left[\mathrm{det}H(\bar{\boldsymbol{k}}) \times \mathrm{det}H(\bar{\boldsymbol{k}}_0)\right],
\label{eq:index}
\end{align}
where $\bar{\boldsymbol{k}}_0$ is the (arbitrarily chosen) origin of the BZ.
We note that $\nu$ is defined modulo $2$, with $\nu =1$ indicating that
the energy gap of the BdG Hamiltonian $H(\bar{\boldsymbol{k}})$
closes at a momentum between $\bar{\boldsymbol{k}}$ and the BZ origin $\bar{\boldsymbol{k}}_0$.
Hence, two-dimensional regions with $\nu=1$ in the BZ are bounded by a gapless line that forms a BFS.
This is demonstrated in Figs.~\ref{fig:figure2}(b)-\ref{fig:figure2}(d), where
we plot $\nu$ in the two-dimensional BZ for   three different parameter choices indicated
in Fig.~\ref{fig:figure2}(a), namely 
$(\lambda/\mu_n,V_s/\Delta,\theta/\pi)=(1.0,0.2,0)$, $(1.0,0.2,0.5\pi)$, and $(1.0,0.3,0)$, respectively.
We observe pairs of thin banana-shaped regions where $\nu =1$, which are bounded by BFSs.

The orientation of these BFSs is controlled by the field angle $\theta$. That is, with increasing $\theta$ the BFSs rotate 
counterclockwise from the 12'o clock and 6'o clock positions for $\theta = 0$ [Fig.~\ref{fig:figure2}(b)] to the
3'o clock and 9'o clock positions for $\theta = \pi / 2$ [Fig.~\ref{fig:figure2}(c)].
The length of the banana-shaped BFSs can be tuned by the Zeeman field  $V_s$.
I.e., with increasing $V_s$ the BFSs become longer and longer. For sufficiently large $V_s$ 
a phase transition point is reached, at which two more point-like BFSs appear, which
are inflated into banana shapes, by further increasing $V_s$.
As a function of Zeeman field $V_s$ and SOC $\lambda$ we 
observe three regions that exhibit BFSs [phases II, III, and IV in Fig.~\ref{fig:figure2}(a)].
Phase II has two BFSs, while phases III and IV have four BFSs.
We note that phases III and IV are separated by a thin line, where the two pairs
of BFSs touch each other, cf.~dashed line in Fig.~\ref{fig:figure2}(a) and 
Figs.~\ref{fig:supplemental1}(b) and~\ref{fig:supplemental1}(e) in the SM. 
In the next section we discuss the conductance and noise spectra
of phases I, II, and III, and the transitions between them.
The  transition between phases III and IV 
is investigated in the SM~\cite{sm_note}.

%------------------------------------------------------------------------------
%------------------------------------------------------------------------------

\section{Differential conductance and noise spectroscopy} \label{sec_IV}

%------------------------------------------------------------------------------
\begin{figure}[t!]
\begin{center}
\includegraphics[width=0.5\textwidth]{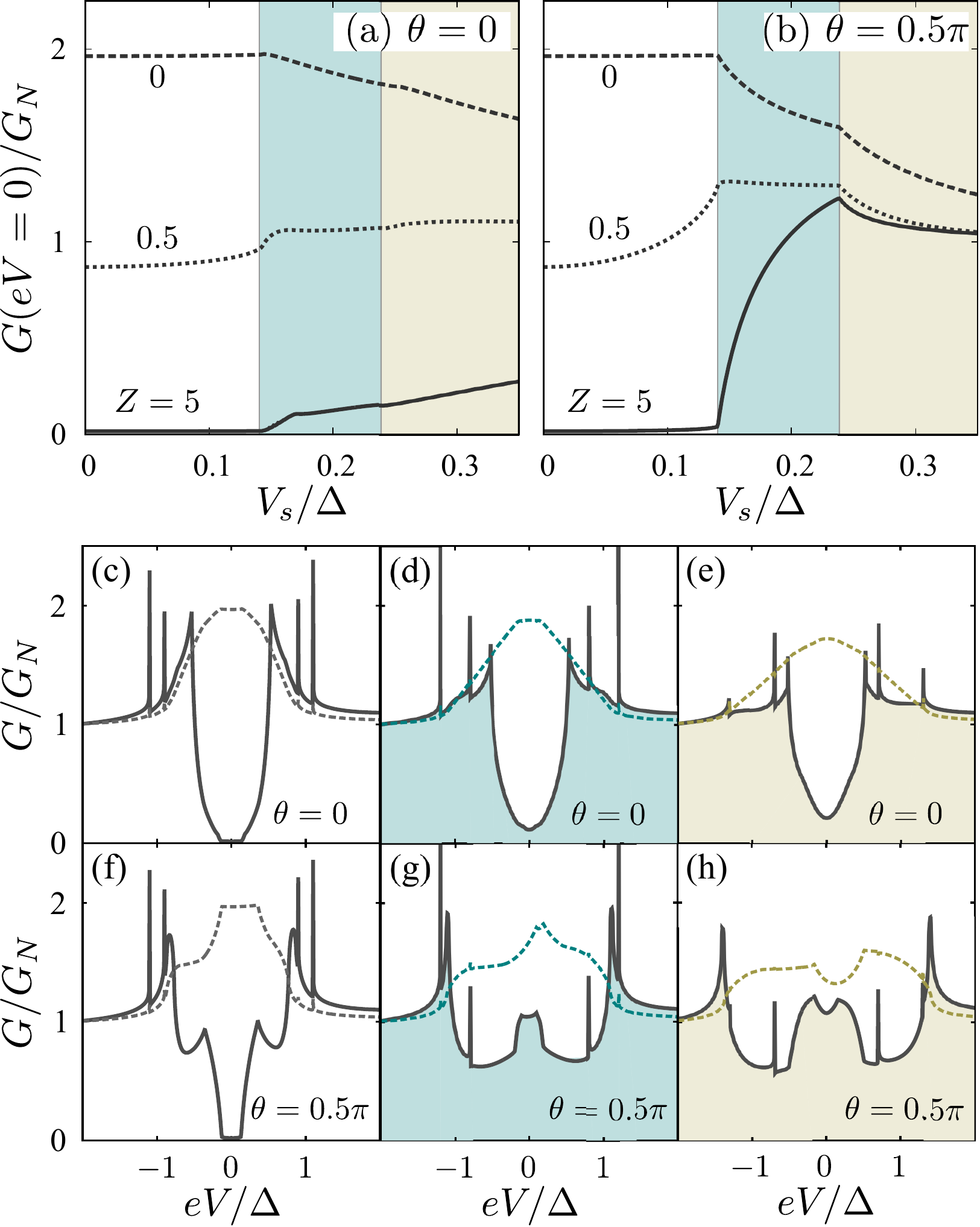}
\caption{Panels (a) and (b) show the 
normalized zero-bias differential conductance $G / G_N$ as a function of   Zeeman field~$V_s$
for the field angles $\theta = 0$ and $\theta =  \pi / 2$, respectively, with
three different junction transparencies $Z$. 
Panels (c)-(h) display the differential conductance as a function of bias voltage
for field angle $\theta=0$ (top row) and $\theta =  \pi / 2$ (bottom row)
and junction transparency $Z=0$ (dashed lines) and $Z=5$ solid lines.
The first column [(c) and (f)] corresponds to phase~I (no BFSs) with $V_s=0.1\Delta$,
the second column [(d) and (g)] represents phase~II (two BFSs) with  $V_s=0.2\Delta$,
while the third column [(e) and (h)] is in phase~III (four BFSs) with $V_s=0.3\Delta$.
}
\label{fig:figure3}
\end{center}
\end{figure}
%------------------------------------------------------------------------------

\begin{figure}[tttt]
\begin{center}
\includegraphics[width=0.5\textwidth]{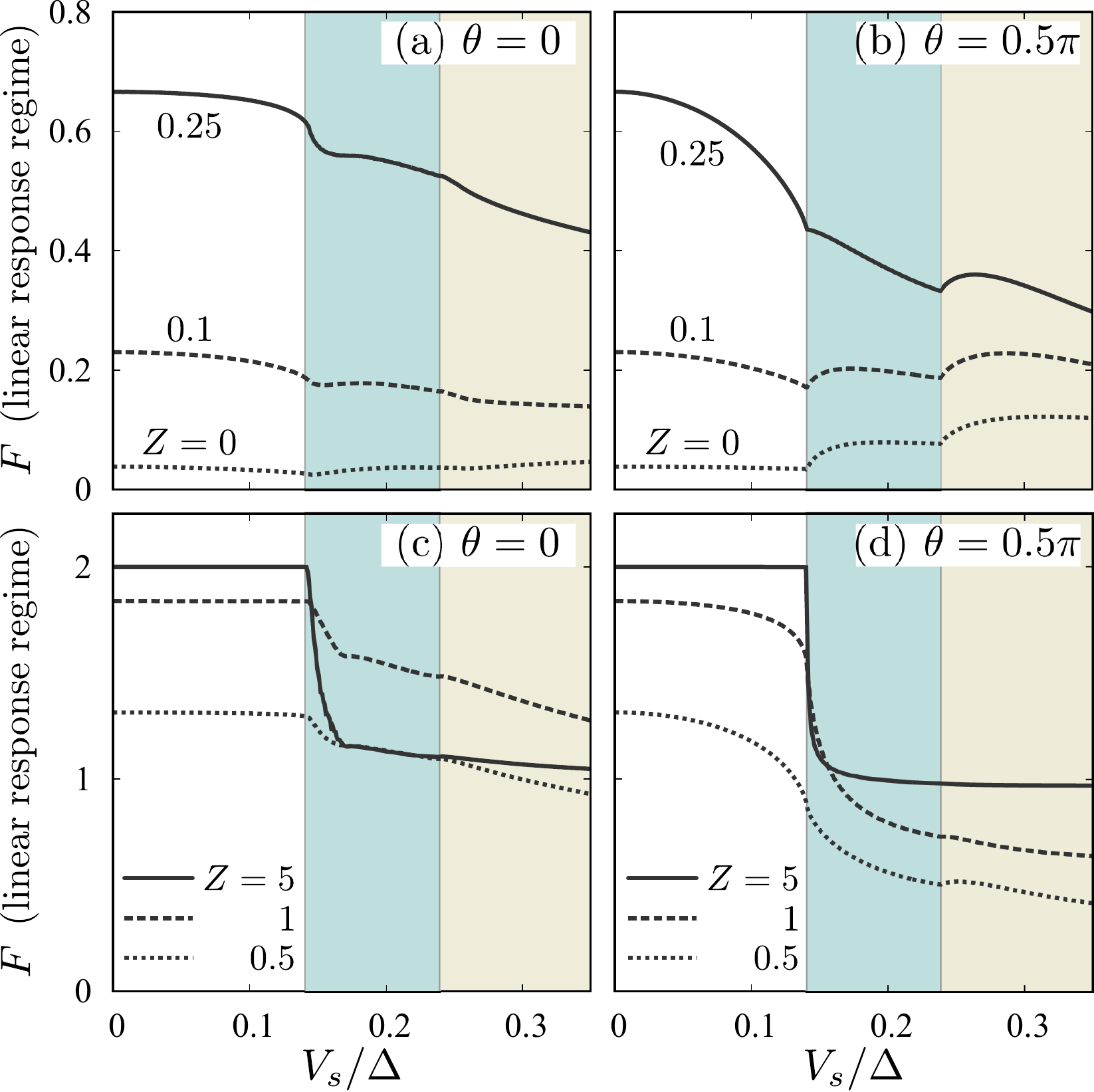}
\caption{
Fano factor $F$ within the linear response regime, Eq.~\eqref{def_Fano_linear}, as a function of Zeeman field $V_s$
for six different junction transparencies $Z$. 
Panels (a) and (c) correspond to field angle $\theta=0$, while in panels (b) and (d)
the field angle is set to $\theta=\pi / 2$.
The white, green, and grey regions correspond to phase I (no BFS), phase II (two BFSs) and phase III (four BFSs),
respectively, see Fig.~\ref{fig:figure2}(a). 
}
\label{fig:figure4}
\end{center}
\end{figure}

Let us now turn to the transport properties, which we use to characterize the
different phases I, II, and III of Fig.~\ref{fig:figure2}. 
In order to compute the differential conductance $G$ and the current-noise ratio $F$, we 
impose periodic boundary conditions in the $y$-direction of the considered device (Fig.~\ref{fig:figure1}).
For concreteness we fix the value of the Rashba SOC to $\lambda / \mu_s =1.0$ and study
the transport properties as a function of Zeeman field $V_s$, with $V_n = 5 V_s$. 
By examining Fig.~\ref{fig:figure2}, we find that for $\lambda/\mu_s=1.0$, phase~I (no BFSs) occurs
for $V_s/\Delta < 0.1405$, phase II (two BFSs) is realized  
for $0.1405<V_s/\Delta<0.2388$, while phase III (four BFSs) exists for $V_s/\Delta>0.2388$.
These three phases are marked in Fig.~\ref{fig:figure3} and Fig.~\ref{fig:figure4} by white, green,
and grey color, respectively. 

We first examine the differential conductance at zero temperature, which we compute within the 
Blonder--Tinkham--Klapwijk formalism (BTK)~\cite{blonder_transition_1982}.
For this purpose we apply a bias voltage $eV$ to our device, which injects
electrons from $i_x = - \infty$.
The differential conductance is evaluated by
\begin{subequations}
\begin{gather}
G(eV)=\frac{e^2}{h}{\sum_{k_y}}^{\prime} \mathrm{Tr} \left[ \hat{\sigma}_0 - \hat{R}_e + \hat{R}_h \right]_{E=eV},\\
\hat{R}_{\alpha} = \hat{r}_{\alpha} \hat{r}_{\alpha}^{\dagger}, \quad
\left( \hat{r}_{\alpha} \right)_{\sigma,\sigma^{\prime}}=r^{\alpha e}_{\sigma \sigma^{\prime}}(k_y,E),
\end{gather}
\end{subequations}
where $r^{\alpha e}_{\sigma \sigma^{\prime}}(k_y,E)$ for $\alpha=e$ ($\alpha = h$) represents the normal (Andreev) reflection coefficient
from an electron with spin $\sigma^{\prime}$ to an electron (hole) with spin $\sigma$ at momentum $k_y$ and energy $E$.
$\sum_{k_y}^{\prime}$ denotes the summation over $k_y$ of the propagating channels.
The reflection coefficients are calculated using lattice Green's function techniques~\cite{lee_anderson_1981,ando_quantum_1991}.  
In the figures we show the normalized differential conductance $G / G_N$, which is obtained by dividing
by the normal conductance $G_N$ calculated by setting $\Delta = eV = V_s = V_n = 0$. 

In Figs.~\ref{fig:figure3}(c)-(h) we show  $G / G_N$ as a function
of bias voltage for the three phases of Fig.~\ref{fig:figure2}(a) and the field angles $\theta = 0$ [top row, panels (c)-(e)] and
$\theta = \pi / 2$ [bottom row, panels (f)-(h)]. 
Figs.~\ref{fig:figure3}(a) and~\ref{fig:figure3}(b) display $G / G_N$ at zero bias as a function
of  Zeeman field $V_s$ for $\theta = 0$ and $\theta = \pi /2$, respectively. 
Let us first focus on the tunneling limit with low junction transparency $Z=5$ (solid lines).
For the tunneling limit we observe in all panels (c)-(h)
two peaks symmetrically located around $\pm \Delta$. These double peaks are the coherence
peaks of the parent superconductor, which are split by the applied Zeeman field (cf.~Fig.~\ref{fig:supplemental2} of the SM~\cite{sm_note}).
In addition there are smaller peaks (mostly at smaller bias), which originate
from the proximity induced superconductivity of the semimetal. 
Around zero-bias voltage we observe clear signatures of the BFSs, i.e., in the tunneling limit  there
is residual conductance at zero bias originating from the finite zero-energy density of states of the BFSs.
The conductance around zero bias increases with increasing $V_s$ and shows
kink singularities at the phase transitions points, where the number of BFSs jumps from zero to two and 
from two to four (see solid lines in Figs.~\ref{fig:figure3}(a) and~\ref{fig:figure3}(b)].
Interestingly,  the zero-bias conductance is larger for $\theta = \pi /2$ than for  $\theta = 0$.
 This is due to the fact that  for $\theta = 0$ most of the incident electrons (i.e., those with $|k_y / \pi | \lesssim 0.05$)
 cannot couple to the BFSs, while for $\theta = \pi / 2$ almost all incident electrons can tunnel into the BFSs [see Figs.~\ref{fig:figure1}(b)-(d)].

Also in the ballistic limit with a completely transparent junction ($Z=0$) distinct signatures of the BFSs can be observed in 
$G / G_N$ around zero bias (dashed lines in Fig.~\ref{fig:figure3}).
That is, in the absence of BFSs all incident electrons show perfect Andreev reflection yielding $G / G_N = 2$, while
in the presence of BFSs only parts of the electrons Andreev reflect leading to a suppression of $G / G_N$, with
 $1 < G/G_N < 2$. As in the tunneling limit,
we observe distinct  kink singularities at the transition points that
separate phases with different numbers of BFSs [see dashed lines in Figs.~\ref{fig:figure3}(a) and~\ref{fig:figure3}(b)].
In passing, we note that the conductance spectra (particularly in the ballistic limit $Z=0$) are not symmetric
around zero bias. The reason for this is the broken time-reversal and broken inversion symmetry, as discussed in more
detail in the SM~\cite{sm_note}.

%------------------------------------------------------------------------------
%------------------------------------------------------------------------------

Let us now turn to the noise spectra of the BFSs. For this purpose we compute
the current-noise ratio (Fano factor)  within the linear response regime, i.e., for $eV \ll \Delta$, 
where the BFSs show a dominant contribution.
The current-noise ratio is given by the ratio
$F = S / e \bar{I}$ between the zero-frequency noise power  
\begin{align}
S = 2 \int^{\infty}_{-\infty} \overline{\delta I(0) \delta I(t)} dt
\end{align}
and the time averaged current at zero temperature $\bar{I}$.
Here, $\delta I(t) = I(t) - \bar{I}$ denotes the deviation of the current at time $t$ from the averaged current $\bar{I}$.
Within the linear response regime, the Fano factor at zero bias simplifies to
\begin{align} \label{def_Fano_linear}
F = \frac{P(eV=0)}{e G(eV=0)},
\end{align}
where the  noise power $P (eV =0) $~\cite{de_jong_doubled_1994,anantram_current_1996}  
is evaluated within the  BTK formalism~\cite{blonder_transition_1982}, as
\begin{equation}
P( 0 ) = \frac{e^3}{h}{\sum_{k_y}}^{\prime} \mathrm{Tr}\left[
\sum_{\alpha=e,h}\hat{R}_{\alpha}(\hat{\sigma}_0-\hat{R}_{\alpha})+ 2\hat{R}_{e}\hat{R}_{h}\right]_{E=0} .
\end{equation}
As before, we compute the reflection coefficients entering in the above formula by use
of lattice Greens function techniques~\cite{lee_anderson_1981,ando_quantum_1991}. 
 
 In Fig.~\ref{fig:figure4} we present the zero-bias Fano factor, Eq.~\eqref{def_Fano_linear}, as a function of Zeeman field $V_s$ for 
 six different junction transparencies $Z$. The left column [(a) and (c)] corresponds to the field angle $\theta=0$,
 while the right column [(b) and (d)] shows the results for the field angle $\theta = \pi/2$.
 The different phases of Fig.~\ref{fig:figure2}(a), with different numbers of BFSs,
 are indicated by the white, green, and grey colors, respectively.
 In the tunneling limit with  low junction transparency $Z=5$ the Fano factor gives us information
 about the fundamental charge of the quasiparticles that tunnel across the junction.
 In phase I without BFSs (white regions), the Fano factor takes the value of two, since 
 the tunneling current is carried exclusively by Cooper pairs composed of two electrons.
 In phases II and III, however, which exhibit a large number of BQPs forming BFSs (green and grey regions), the Fano factor decreases to a value substantially lower than two.
This indicates that the current is carried by both Cooper pairs and BQPs, 
which have an effective charge substantially smaller than that of Cooper pairs. 
Remarkably, if we set the field angle to $\theta = \pi /2$, such that almost all incident electrons can couple to the BFSs,  
the Fano factor  drops below one [solid line in Fig.~\ref{fig:figure4}(c)], implying that almost all of the current is carried by the BQPs.
 
In the ballistic limit with transparent junction $Z=0$, we can obtain information about the types of scattering processes that occur at the junction barrier.
In phase~I without BFSs we observe that the Fano factor is almost zero [dotted lines in Figs.~\ref{fig:figure4}(a) and~\ref{fig:figure4}(b)], as almost all incident electrons undergo 
Andreev reflection such that the noise power $P$ is almost zero.  
In phases II and III with BFSs, however, the Fano factor increases, indicating that some electrons are transformed into BQPs of the BFSs, while others Andreev reflect.
By setting the field angle to $\theta = \pi /2$, we can maximize the coupling between the lead electrons and the BFS, such
that more electrons transmit into the BFSs, thereby increasing the Fano factor [cf.~dotted lines in Figs.~\ref{fig:figure4}(a) and~\ref{fig:figure4}(b)]. 
As in Fig.~\ref{fig:figure3} we observe distinct kinks in the Fano factor at the transition points that separate the different phases.

\section{Conclusions}

In summary, we have investigated the differential conductance and noise spectroscopy of a NS junction with BFSs. 
We find that the current-noise ratio (Fano factor) provides key fingerprints of the presence of BFSs (see Fig.~\ref{fig:figure4}): In the tunneling limit
the Fano factor drops from two to about one, whenever there are BFSs present. In the ballistic limit,
on the other hand, the Fano factor jumps from zero to a finite value, due to the presence of BFSs.
Moroever, the Fano factor as a function of Zeeman field exhibits 
kink discontinuities at the transition lines that separate phases with different number of BFSs. 
We hope that these predictions will help experimentalists to identify the presence of BFSs in their superconducting hybrid devices. 

We note that the creation of BFSs in semiconductor/superconductor hybrid devices requires Zeeman fields larger 
than the proximity-induced pair potential in the semiconductor, but smaller than the critical field of the parent superconductor.
 This condition is realized for hybrid devices, such as Al/InAs heterostructures \cite{phan_detecting_2022,marcus_nat_commun_2016}, 
 where the $g$ factor of the  semiconductor is large, while the proximity-induced pair potential is small compared
 to the parent superconductor.
While we have focused in this Letter on superconducting hybrids with BFSs, the obtained results are of relevance also 
to other systems with BFSs, for example, multi-band superconductors with broken time-reversal symmetry~\cite{agterberg_bogoliubov_2017-1,timm_inflated_2017-1,brydon_bogoliubov_2018-1,menke_bogoliubov_2019}.

In closing, we mention several promising directions for future research. First, it would be
interesting to investigate other types of junctions, e.g., NSN and SNS junctions, involving superconductors
with BFSs, and to determine the signatures of BFSs in the Josephson current. Experimentally, it should be straightforward to fabricate NSN and SNS junctions using existing
thin film technology. Second, it would be of interest to study thermal transport properties of BFSs. 
While conventional superconductors are prefect thermal insulators at low temperature, superconductors with BFSs are expected to 
show substantial thermal transport carried by the large density of BQPs near zero energy.
Third, the interplay between BFSs and Andreev surface bound states (or interface bound states) and its effects on the transport properties
would be a worthwhile topic for future research.

\acknowledgements
The authors thank O.~Maistrenko and P.~Bonetti for useful discussions.
S.I. is supported by the Grant-in-Aid for JSPS Fellows (JSPS KAKENHI Grant No. JP21J00041) and the JSPS Core-to-Core Program (No. JPJSCCA20170002).
S.B.  thanks the MPI-FKF for financial support.

\bibliographystyle{apsrev4-1}
\bibliography{bfs_literature}

\clearpage

\pagebreak

\setcounter{figure}{0}
\makeatletter 
\renewcommand{\thefigure}{S\@arabic\c@figure} 

\setcounter{equation}{0}
\makeatletter 
\renewcommand{\theequation}{S\@arabic\c@equation} 

\setcounter{section}{0}
\makeatletter 
\renewcommand{\thesection}{S\@arabic\c@section} 

\makeatother

\widetext
\begin{center}
{\Large{\bf Supplemental Material for} \\
\vspace{0.6cm}
{\large ``{\bf Anomalous Fano factor as a signature of Bogoliubov Fermi surfaces}''}
}

\vspace{0.6cm}

\textbf{Authors:}  Sayan Banerjee, Satoshi Ikegaya, and Andreas P. Schnyder

\end{center}

In this supplemental material we provide the expression for the momentum space Hamiltonian $H(\bar{\boldsymbol{k}})$
in Eq.~\eqref{eq_phase_diag_ham} of the main text (Sec.~\ref{sm_sec_eins})
and discuss the transition between phases III and IV (Sec.~\ref{sm_sec_zwei}).
In addition we present in Sec.~\ref{sec_SM_add_conductance_figs}
two additional figures of the differential conductance $G$, one without Zeeman field, and
one comparing the two field angles $\theta=-0.5\pi$ and $\theta=0.5\pi$.

\section{Momentum space Hamiltonian $H(\bar{\boldsymbol{k}})$}
\label{sm_sec_eins}

The momentum space Hamiltonian $H(\bar{\boldsymbol{k}})$ in Eq.~\eqref{eq_phase_diag_ham} of the main text is given by
\begin{align} \label{momentum_Ham_SM}
&H(\bar{\boldsymbol{k}})  = \left[ \begin{array}{cccccc}
\check{H}_n(\bar{\boldsymbol{k}}) & \check{T}_i & & & & \mbox{\Large 0} \\
\check{T}_i & \check{H}_s(\bar{\boldsymbol{k}}) & \check{T}_s & & \\
& \check{T}_s & \check{H}_s(\bar{\boldsymbol{k}}) & \check{T}_s & &\\
& & \ddots& \ddots & \ddots & \\
&  & & \check{T}_s & \check{H}_s(\bar{\boldsymbol{k}}) & \check{T}_s \\
\mbox{\Large 0} & & & & \check{T}_s & \check{H}_s(\bar{\boldsymbol{k}})\\
\end{array}\right],\\
\nonumber
&\check{H}_n(\bar{\boldsymbol{k}})  = \left[ \begin{array}{cc} \hat{h}_n(\bar{\boldsymbol{k}}) & 0 \\ 0 & -\hat{h}^{\ast}_n(-\bar{\boldsymbol{k}})\end{array}\right],\quad
\check{H}_s(\bar{\boldsymbol{k}})  = \left[ \begin{array}{cc} \hat{h}_s(\bar{\boldsymbol{k}}) & \Delta (i \hat{\sigma}_y) \\
-\Delta (i \hat{\sigma}_y) & -\hat{h}^{\ast}_s(-\bar{\boldsymbol{k}})\end{array}\right],\quad
\check{T}_{i(s)} = \left[ \begin{array}{cc} -t_{i(s)} \hat{\sigma}_0 & 0 \\
0 & t_{i(s)} \hat{\sigma}_0 \end{array}\right],\\
\nonumber
&\hat{h}_n(\bar{\boldsymbol{k}})  = \left[ \begin{array}{cc} 
\xi_n(\bar{\boldsymbol{k}}) & \lambda(\bar{\boldsymbol{k}}) + V_n e^{-i \theta}\\
\lambda^{\ast}(\bar{\boldsymbol{k}}) + V_n e^{i \theta} & \xi_n(\bar{\boldsymbol{k}}) \end{array}\right],\quad
\hat{h}_s(\bar{\boldsymbol{k}})  = \left[ \begin{array}{cc} 
\xi_s(\bar{\boldsymbol{k}}) &  V_s e^{-i \theta}\\
\nonumber
V_s e^{i \theta} & \xi_s(\bar{\boldsymbol{k}}) \end{array}\right],\\
&\xi_{n(s)}(\bar{\boldsymbol{k}}) = -2t_{n(s)} ( \cos k_x + \cos k_y)  - \epsilon_{n(s)}, \quad
\lambda(\bar{\boldsymbol{k}}) = \lambda(\sin k_y + i \sin k_x),
\nonumber
\end{align}
where $t_{n(s)}$ denotes the hopping amplitude in the semiconductor (superconductor),
and $\epsilon_n=4t_n-\mu_n$ ($\epsilon_s=6t_s-\mu_s$) with $\mu_{n(s)}$ representing the chemical potential in the semiconductor (superconductor),
and the Zeeman potential in the semiconductor (superconductor) is given by $\boldsymbol{V}_{n(s)}=V_{n(s)}(\cos \theta, \sin \theta, 0)$.
The strength of the Rashba spin-orbit coupling potential in the semiconductor is represented by $\lambda$,
and the pair potential in the superconductor is denoted by $\Delta$.
The hopping integral between the semiconductor and the superconductor is given by $t_i$.
The Pauli matrices in spin space are given by $\hat{\boldsymbol{\sigma}}=(\hat{\sigma}_x,\hat{\sigma}_y,\hat{\sigma}_z)$, and $\hat{\sigma}_0$ denotes the $(2\times2)$ unit matrix.

\section{Transition between phases III and~IV of Fig.~\ref{fig:figure2}}
\label{sm_sec_zwei}

%------------------------------------------------------------------------------
\begin{figure}[t!]
\begin{center}
\includegraphics[width=0.5\textwidth]{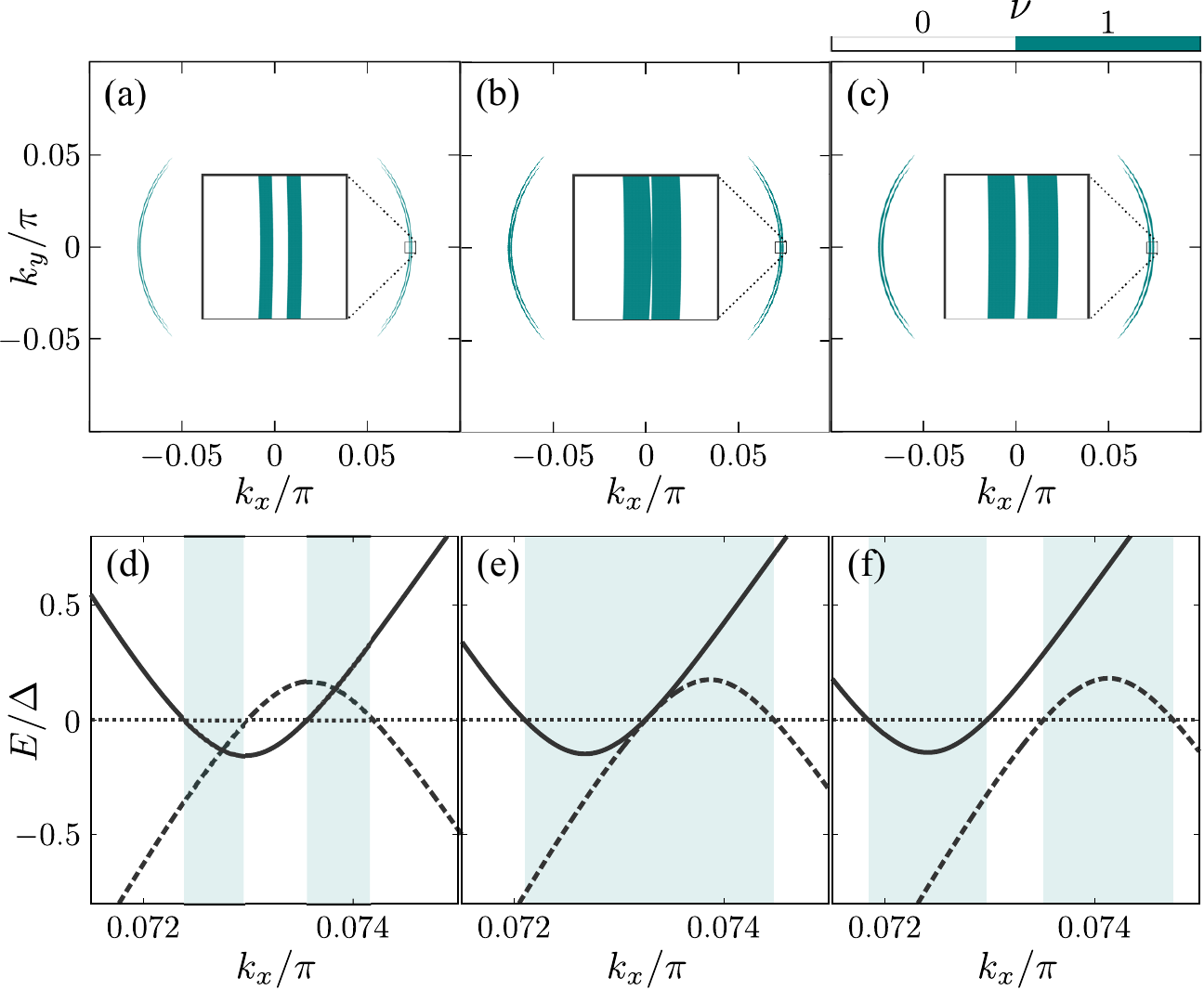}
\caption{Panels (a), (b), and (c) show the determinant index~$\nu$, Eq.~(\ref{eq:index}), in the two-dimensional Brillouin zone
for $\lambda / \mu_n = 0.04$, $0.083$, and $0.12$, respectively. The Zeeman field and the field angle 
are fixed to $V_s / \Delta = 0.24$ and $\theta =0$.
The colored regions are bounded by BFSs.
Panels (d), (e), and (f) show the energy dispersion of $H(\bar{\boldsymbol{k}})$, Eq.~\eqref{momentum_Ham_SM},
for the same parameters as in (a), (b), and (c), respectively, 
with $k_y=0$ and as a function of $k_x$.
Note that panels (a)/(d) and (c)/(f) correspond to phases IV and III, respectively,
while panels (b)/(e) are at the phase boundary, cf.~Fig.~\ref{fig:figure1}(a) of the main text.
}
\label{fig:supplemental1}
\end{center}
\end{figure}
%------------------------------------------------------------------------------

In this section, we study the phase transition between the phases III and  IV in Fig.~\ref{fig:figure2}(a) of the main text.
In what follows, we fix $V_s/\Delta = 0.24$ and $\theta =0$, and vary the strength of the Rashba SOC $\lambda$.
Other parameters are chosen to be the same as those in the main text. 
The phase boundary between the phases III and  IV is located at $\lambda/\mu_n \simeq 0.083$.
In Figs.~\ref{fig:supplemental1}(a), (b), and (c), we show the determinant index $\nu$, Eq.~\eqref{eq:index},  in the two-dimensional Brillouin zone
for the parameters $\lambda/\mu_n=0.04$ (phase IV), 0.083 (phase boundary), and 0.12 (phase III), respectively.
We observe that with increasing $\lambda/\mu_n$, the two pairs 
of BFSs approach each  other, until they merge at  $\lambda/\mu_n \simeq 0.083$, 
and then separate again.

In Figs.~\ref{fig:supplemental1}(d), (e), and (f), we show the energy dispersion $E(\bar{\boldsymbol{k}})$ 
of the momentum space Hamiltonian $H(\bar{\boldsymbol{k}})$, Eq.~\eqref{momentum_Ham_SM},  
for the same parameters as in Figs.~\ref{fig:supplemental1}(a), (b), and (c), respectively,
for fixed $k_y=0$ and as a function of $k_x$. Here, we focus on the pair of BFSs at 3'o clock, i.e., near $k_x=0.074\pi$.
We observe two parabolas, one with positive curvature and one with negative curvature, whose crossing points with the Fermi level define
the BFSs. This is indicated by the shaded regions, which correspond to the momenta inside the BFSs.
Interestingly, in phase~IV the positive-curvature and negative-curvature parabolas intersect each other  [panel (a)],
while in phase~III they avoid each other  [panel (f)].
At the phase transition the two parabolas touch each other [panel (a)], such that the
two pairs of BFSs merge together. 
Hence, the transition from phase III to phase IV represent a Lifshitz-type transition, where
the topology of the BFSs changes.

\section{Additional figures of the differential conductance} \label{sec_SM_add_conductance_figs}

\begin{figure}[hhhh]
\begin{center}
\includegraphics[width=0.475\textwidth]{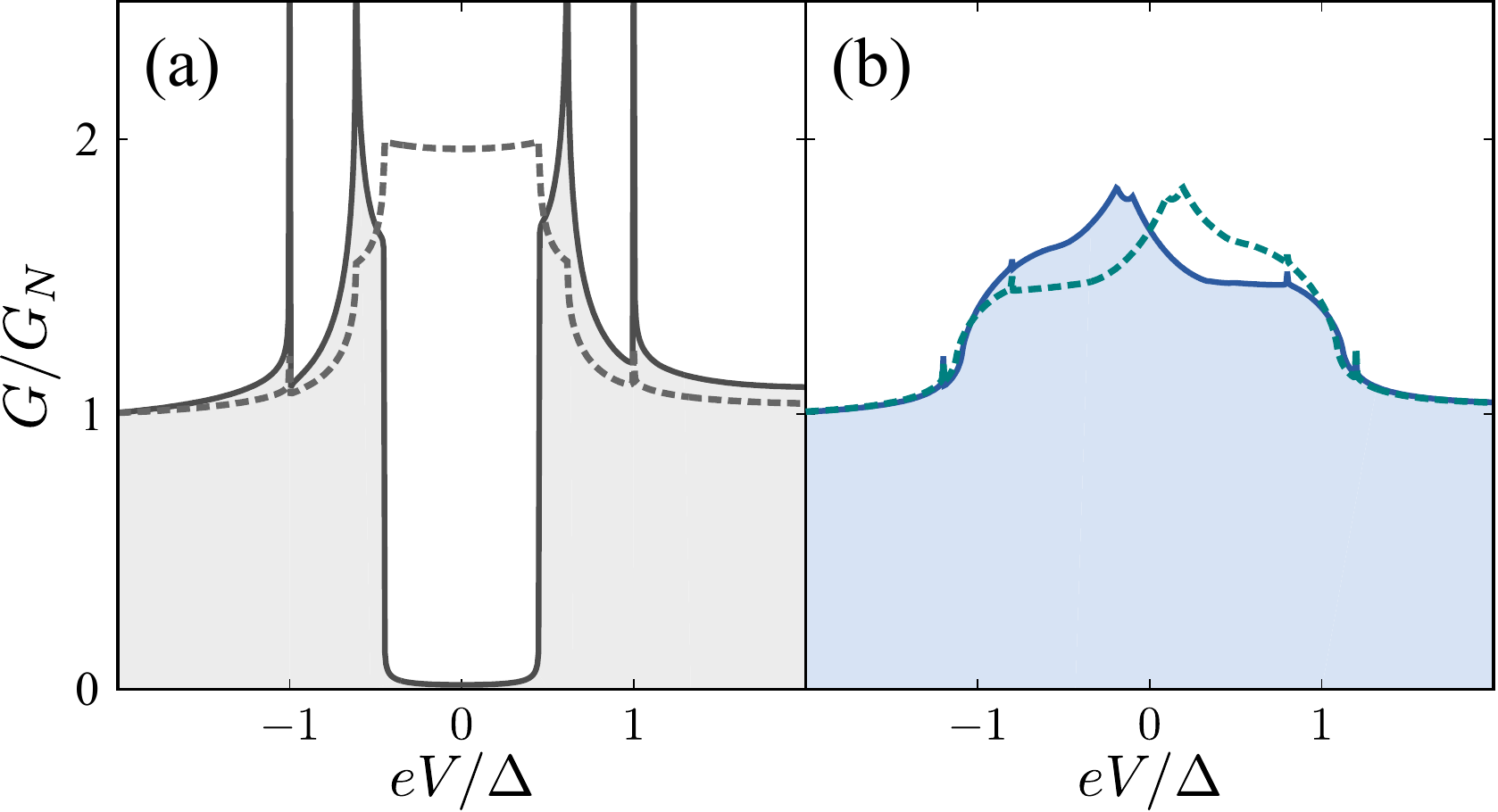}
\caption{
(a) 
Differential conductance in the absence of a Zeeman field, i.e., $V_s = V_n =0$, for 
$\lambda/\mu_s = 1.0$. The solid and dashed lines represent the tunneling limit ($Z=5$) and ballistic limit ($Z=0$), respectively.
(b) 
Differential conductance for oppositely oriented fields for $V_s = 0.2 \Delta$, $\lambda/\mu_s = 1.0$, and $Z=0$. 
The solid and dashed lines show to the results for $\theta=-0.5\pi$ and $\theta=0.5\pi$, respectively.
}
\label{fig:supplemental2}
\end{center}
\end{figure}

In Fig.~\ref{fig:supplemental2}(a) we show the normalized differential conductance $G / G_N$ as a function of   bias voltage
in the absence of a Zeeman field, i.e., for $V_s = V_n =0$. In this case the spectra of the BdG Hamiltonian
are symmetric with respect to zero energy for any fixed $k_y$, due to time-reversal and particle-hole symmetry.
This leads to completely symmetric differential conductance spectra, i.e., $G / G_N$ does not change upon reversing the bias voltage. 
In the device of Fig.~\ref{fig:figure1}  the parent superconductor with 
pairing gap $\Delta$ induces a smaller gap $\Delta_{\textrm{ind}}$ in the semiconductor, i.e., $\Delta_{\textrm{ind}} < \Delta$.
This leads to three different voltage regimes with different conductance characteristics. Let us first discuss this for the
case of perfect junction transparency $Z=0$ (dashed line):
(i) For bias voltage larger than $\Delta$ there is no Andreev reflection possible leading to 
$G / G_N = 1$,
(ii) for bias voltage smaller than $\Delta$ but larger than $\Delta_{\textrm{ind}}$, Andreev reflection 
occurs only for electrons with suitable momenta, mostly via the parent superconductor, leading to $1< G / G_N < 2$, and
(iii) for bias voltage smaller then $\Delta_{\textrm{ind}}$, perfect Andreev reflection is possible 
for all incident electrons, yielding $G / G_N = 2$.
Accordingly, we observe in the tunneling limit $Z=5$ (solid line) sharp peaks 
at the boundary of these regimes.

In Fig.~\ref{fig:supplemental2}(b) we show the normalized differential conductance $G/G_N$
for opposite field orientations, i.e., for $\theta = -0.5 \pi$ and $\theta = +0.5 \pi$ with 
$V_s = 0.2 \Delta$ (see solid and dashed lines, respectively). 
In the presence of a Zeeman field $V_s$ with $\theta \ne 0$ the BdG spectra acquire an energy shift which scales linearly with $k_y$ and $V_s$ with an opposite sign for the
two BdG bands~\cite{yuan_zeeman-induced_2018}. Thus, the BdG spectra are no longer symmetric with respect to zero-energy for fixed $k_y$, in line with the broken time-reversal symmetry.
As a consequence, the three voltage regimes discussed above shift in an asymmetric fashion with respect to zero energy, as can be seen in Fig.~\ref{fig:supplemental2}(b) 
and Figs.~\ref{fig:figure3}(f)-(h).
Note that upon flipping the sign of the Zeeman potential (i.e., by letting $\theta \to \theta + \pi$) the energy shifts of the BdG spectra change sign, 
leading to a reflection of the differential conductance spectra with respect to zero-energy (compare solid and dashed lines).

%------------------------------------------------------------------------------
%------------------------------------------------------------------------------

\clearpage

\end{document}